\documentclass{moriond}

\usepackage{lineno} 

\def\Journal#1#2#3#4{{#1} {\bf #2}, #3 (#4)}

\def\NIMA{{\em Nucl. Instrum. Methods} A}

\def\PLB{{\em Phys. Lett.}  B}
\def\PRL{\em Phys. Rev. Lett.}
\def\PRD{{\em Phys. Rev.} D}

\def\be{\begin{equation}}
\def\ee{\end{equation}}
\def\bea{\begin{eqnarray}}
\def\eea{\end{eqnarray}}

\def\Bs{B_{s}^{0}}
\def\Bd{B^{0}}
\def\jpsi{J/\psi}
\def\phis{\phi_{s}}
\def\dgs{\Delta\Gamma_s}

\def\Bsbar{\bar{B}_{s}^{0}}
\def\jpsipipi{J/\psi\,\pi^+ \pi^-}
\def\bsjpsipipi{B_s^0 \to J/\psi\,\pi^+ \pi^-}
\def\jpsikk{J/\psi\,K^+ K^-}
\def\bsjpsikk{B_s^0 \to J/\psi\,K^+ K^-}
\def\bdjpsikstar{B^0 \to J/\psi\,K^{*0}(\to K^+ \pi^-)}

\newcommand{\Photo}{\includegraphics[height=35mm]{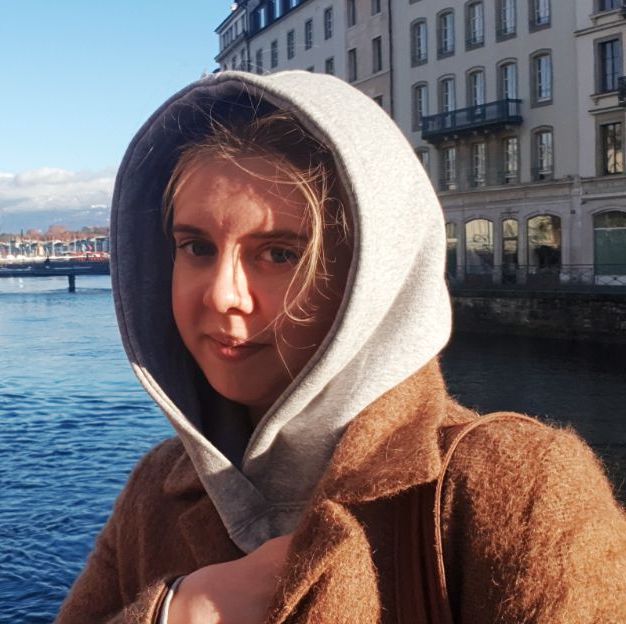}}

\begin{document}
\vspace*{4cm}
\title{MIXING AND TIME-DEPENDENT CP VIOLATION IN BEAUTY AT LHCB}

\author{E. GOVORKOVA on behalf of the LHCb Collaboration}

\address{Nikhef National Institute for Subatomic Physics,\\
Science Park 105, Amsterdam, The Netherlands}

\maketitle\abstracts{Recent measurements of the time-dependent CP
violation are presented. The decays of $\Bs$ mesons to $\jpsikk$
and $\jpsipipi$ final states are used to measure CP-violating parameters
with proton-proton collision data, corresponding to an integrated luminosity
of~1.9~fb$^{-1}$, collected by the LHCb detector at a centre-of-mass energy
of~13~TeV in 2015 and 2016.
}

\section{CP violation in the $\Bs-\Bsbar$ system}
The LHCb detector~\cite{lhcb1,lhcb2} is a general purpose detector in the forward
region, situated at the Large Hadron Collider at CERN. The broad physics program
of the LHCb experiment covers various topics including study of the CP
violation~(CPV) in $B$ mesons. The CP-violating phase $\phi_s$ arises in
the interference between the amplitudes of a $B_s^0$ meson \footnote{The inclusion
of charge-conjugate processes is implied throughout this manuscript, unless
otherwise noted.} decaying via $b \to c\overline{c}s$ transition directly to its
final state or after oscillation to a $\overline{B}$ meson. In the Standard
Model~(SM) ignoring subleading penguin contributions, this phase is constrained
via global fits to the experimental data to be~\mbox{$-0.0364 \pm 0.0016\,\rm{rad}$}~\cite{CKMfitter}.
This precise prediction makes the measurement of $\phis$ interesting since it is possible
that new physics processes could modify the phase, if new particles were to contribute
to the $\Bs-\Bsbar$ mixing diagrams~\cite{Buras:2009if,Chiang:2009ev}. The value
of $\phis$ measured to be significantly different from the SM would be a clear
evidence for physics beyond the SM.

\section{Measurement of $\phis$ at LHCb}
Two recent measurements of the phase $\phis$ at the LHCb experiment are presented.
Both analyses measure the CP-violating phase $\phis$ using $\Bs$ meson
decays. One analysis explores the $\bsjpsikk$ decay mode~\cite{kk},
while the other studies the $\bsjpsipipi$ decay~\cite{pipi}. Since the analysis strategies
are very similar, the baseline strategy adopted by both is covered and relevant
differences between the two are highlighted.

Both analyses use proton-proton collision data collected with the LHCb detector in
2015 and 2016, corresponding to the total integrated luminosity of 1.9~fb$^{-1}$.
A $\Bs$~($\Bsbar$) meson after being produced in a proton-proton collision in the
LHCb detector (so-called primary vertex), flies approximately 1~cm before decaying
inside the vertex locator, VELO, of the LHCb detector. The excellent resolution of
the VELO detector, which is equal to 45.5~(41.5)~fs for the $\bsjpsikk$~($\bsjpsipipi$)
mode, allows to resolve oscillations
in the $\Bs-\Bsbar$ system. The decay products of a $\Bs$~($\Bsbar$) meson,
$\jpsikk$ or $\jpsipipi$ where $\jpsi$ decays to $\mu^+\mu^-$ pair, fly from the
decay point of $\Bs$~(so-called secondary vertex) through the rest of the detector,
traversing the magnet, where tracks of charge particles are bent; tracking
stations; Cherenkov detectors, which allows to identify hadron type; calorimeter
system and muon stations. The distance between the secondary and primary vertices is
translated to the $\Bs$ meson decay-time using estimate of its momentum.

CP-violating parameters $\phis$ and $|\lambda|$ are measured. The $\bsjpsikk$
channel allows measuring lifetime parameters of the $\Bs$ meson: the decay-width
difference between the heavy and light $\Bs$ meson eigenstates, $\dgs = \Gamma_L - \Gamma_H$,
and the average decay-width of the states, $\Gamma_s = \frac{\Gamma_L + \Gamma_H}{2}$.
However, since $\jpsipipi$ final
state is almost entirely CP-odd, only decays of the heavy $\Bs$ meson eigenstate
are possible, therefore one can measure $\Gamma_{H}$ with $\bsjpsipipi$. In order
to disentangle CP-even from CP-odd component, angular analysis is required. For
the four-body final state, three independent angles are needed to describe the
system. Both analyses make use of helicity angles formalism~\cite{helicity}.
In the measurements the decay-width difference between the $\Bs$~($B_{H}$) and
$\Bd$ meson is fitted for, in order to keep results independent of the value of
$\Bd$ meson width. In this way, the value of the $\Gamma_{s/H}$ can be extracted
using the latest world average of the value of $\Gamma_{B^0}$.

The experimental differential decay-time rate for an initial $\Bs$ meson as a
function of decay time and angles is given as~\cite{Liu:2013nea}
\begin{equation}
 \frac{d^4 \Gamma}{dt\,d\Omega} \propto
            \sum_{k=1}^{10} \varepsilon(t, \Omega) f_k(\Omega) h_k(t)
            \otimes G(t|\sigma_t),
 \label{eq:diff_decay_rate}
 \end{equation}
where $\varepsilon(t, \Omega)$ is efficiency as a function of decay-time and
angular observables, $f_k(\Omega)$ are angular functions, $G(t|\sigma_t)$ is
experimental decay-time resolution
and the decay-time-dependent functions $h_k(t)$ for the decay of $\Bs$ meson
produced as $\Bs$ meson are given as
\begin{equation}
 h_k(t)= \frac{3}{ 4\pi}
   e^{-\Gamma t}\left\{ a_k \cosh\frac{\Delta \Gamma t}{2} +
                        b_k \sinh\frac{\Delta \Gamma t}{2} +
                        c_k \cos(\Delta mt)  +
                        d_k \sin(\Delta mt) \right\}.
\label{eq:diff_decay_rate_time}
 \end{equation}

\noindent For an initial $\Bsbar$ at production, the signs of $c_k$ and $d_k$
should be reversed. The end goal is to perform a simultaneous maximum likelihood
fit to the decay-time and three angles in order to extract CP-violating parameters.
As can be seen from the equation~\ref{eq:diff_decay_rate}, there are several
experimental inputs that are required for the time-dependent fit to be performed:
efficiency, decay-time resolution of the detector and the knowledge of the flavour
of the $\Bs$ meson at production. The details on these inputs are given in the
following.

\subsection{Selection and mass fit}
\label{subsec:selection}
The time-dependent angular fit is performed on background-subtracted data. For
both decays under study a corresponding boosted decision tree~\cite{bdt}, BDT,
is used to select signal and reject background candidates. The BDTs are trained
using data sidebands as background proxy and simulated events as signal proxy.
After the training, the optimal cut is found and applied on the data sample.

In case of $\Lambda_{b}\to J/\psi\,p^+ K^-$ decays, it can happen that the proton
in the final state is mis-identified as a kaon and resulting $\jpsikk$ wrong
combination might end up peaking under the signal peak. Since this contribution
is significant for the $\bsjpsikk$ mode it is subtracted by injected negatively
weighted simulated sample of $\Lambda_{b}\to J/\psi p^+ K^-$
decays with the total weight equal to the expected contribution of the
$\Lambda_{b}$ background. Other peaking background contributions coming from
decays $B^{0}\to J/\psi K^+ K^-$, $B^{0}\to J/\psi \pi^+ \pi^-$ and
$B^{0}\to J/\psi K^{*0}(\to K^+ \pi^-)$ are either vetoed using particle identification
requirements or accounted for directly in the mass fit.

\begin{figure}
\begin{minipage}{0.5\linewidth}
\centerline{\includegraphics[width=0.9\linewidth]{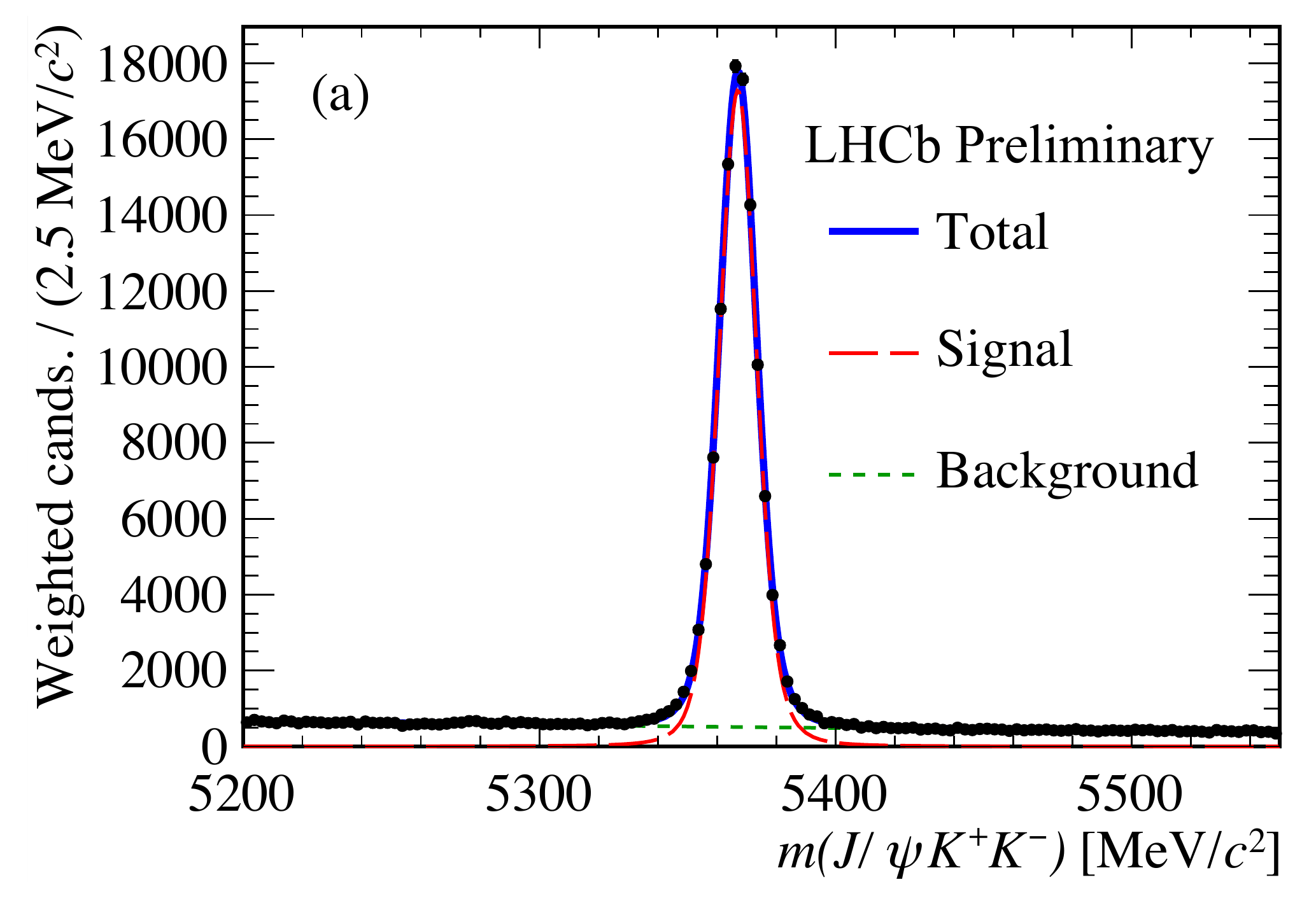}}
\end{minipage}
\hfill
\begin{minipage}{0.5\linewidth}
\centerline{\includegraphics[width=0.9\linewidth]{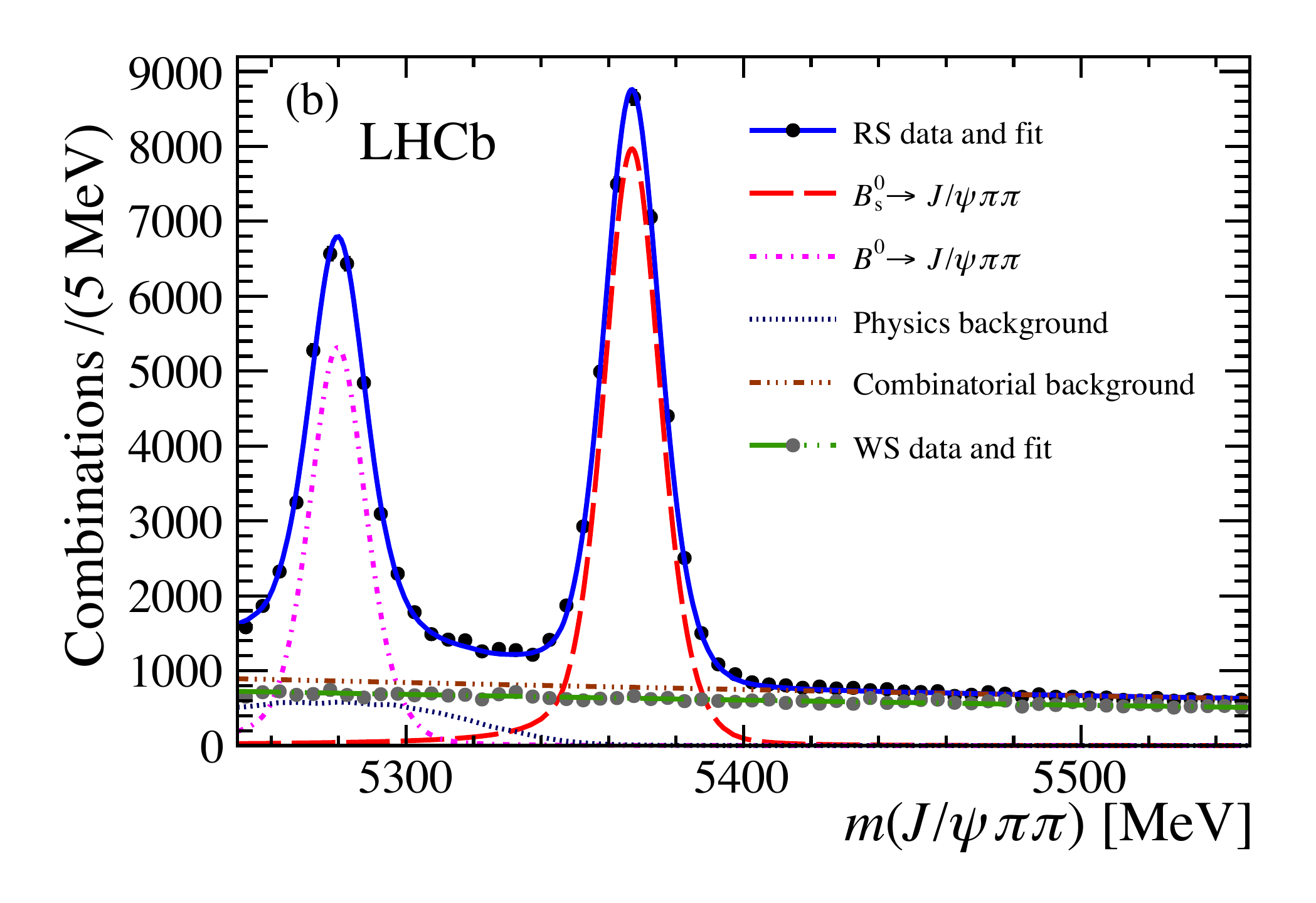}}
\end{minipage}
\caption{Distribution of the invariant mass of selected (a) $\bsjpsikk$~\protect\cite{kk}
    and (b) $\bsjpsipipi$~\protect\cite{pipi} decays. The color-coding of PDFs is explained on the each plot.}
\label{fig:mass_fit}
\end{figure}

In order to disentangle combinatorial background events from signal candidates the
signal weights are used. Those weights are obtained using the $s$Plot procedure~\cite{splot},
which assigns a weight to each event based on the probability density function (PDF)
that is used to describe the invariant mass spectra and are later used to
statistically subtract the background contribution. The invariant mass distributions
of $\jpsikk$ and $\jpsipipi$ are shown in Fig.~\ref{fig:mass_fit} together with
the analytical shapes that are used to extract signal weights.

\subsection{Decay time resolution}\label{subsec:time_resolution}
\begin{figure}[htb]
\begin{minipage}{0.5\linewidth}
\centerline{\includegraphics[width=0.95\linewidth]{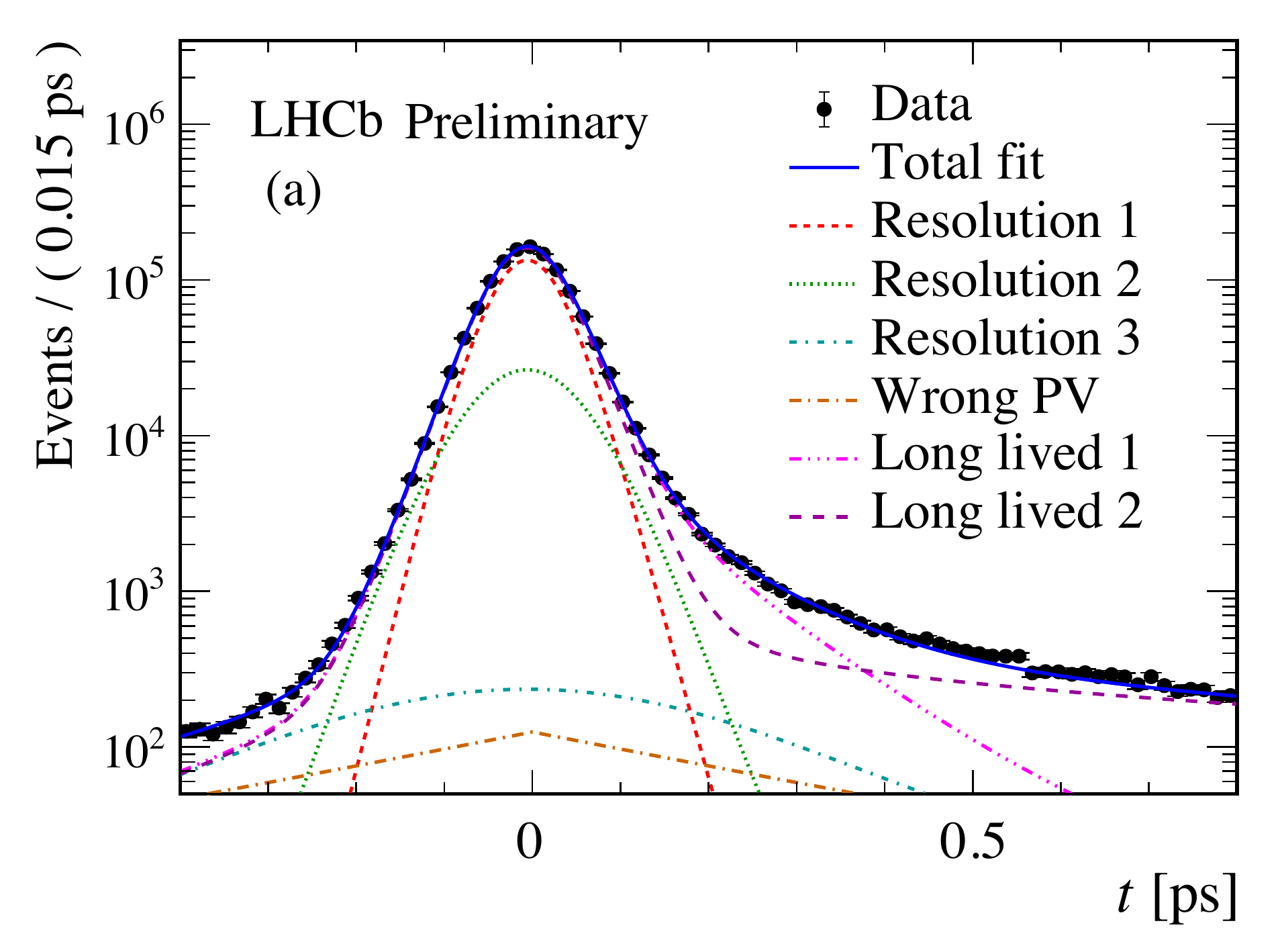}}
\end{minipage}
\hfill
\begin{minipage}{0.52\linewidth}
\centerline{\includegraphics[width=0.98\linewidth]{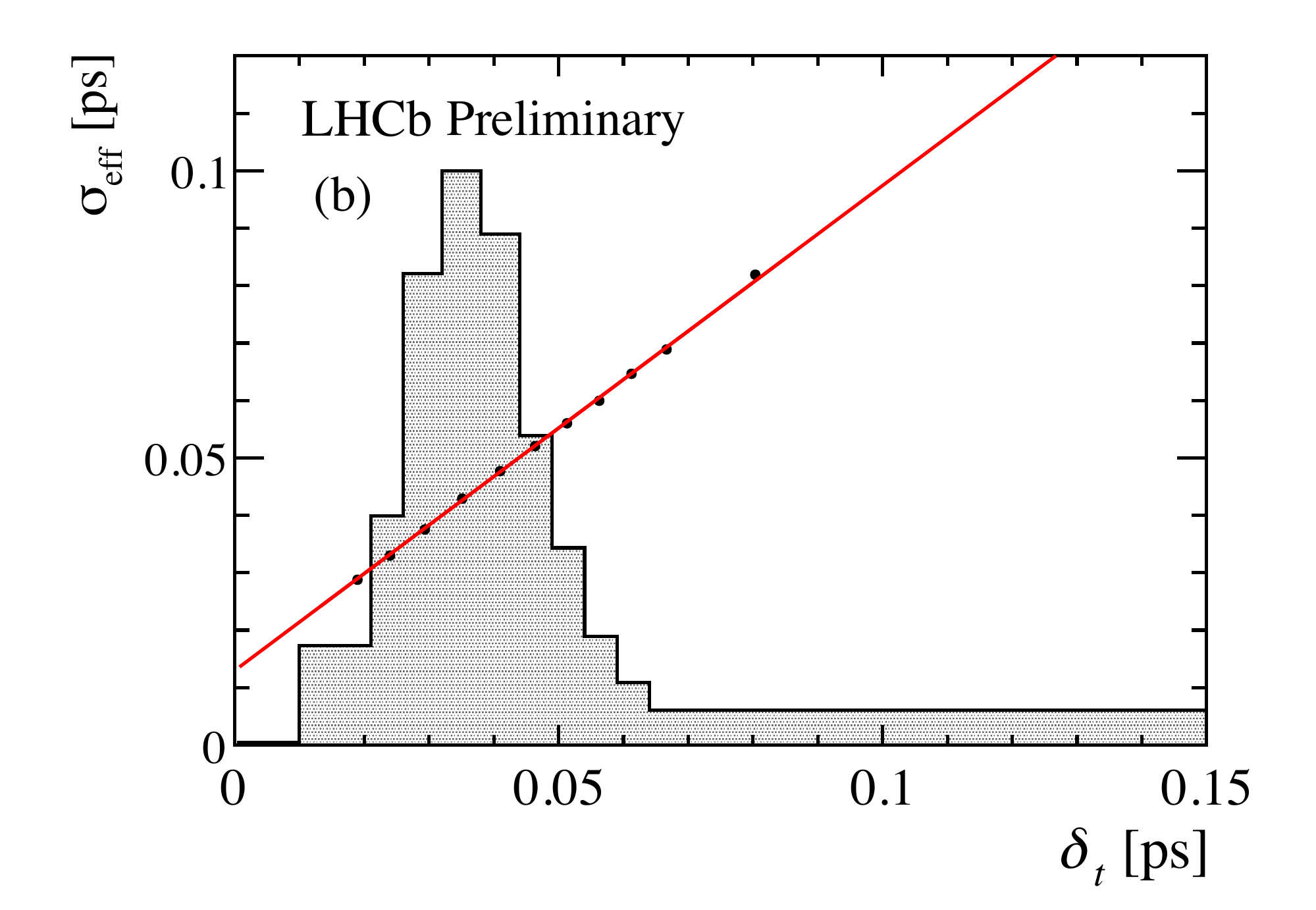}}
\end{minipage}
\caption{(a) Decay-time distribution of the prompt $\bsjpsikk$ calibration
    sample~\protect\cite{kk} with the
    result of an unbinned maximum likelihood fit overlaid in blue. The overall
    resolution is represented by the dashed red line. (b) Variation of the
    effective decay-time resolution, $\sigma_{\rm eff}$, as a function of the
    estimated per-candidate decay-time uncertainty, $\delta_t$. The red line
    shows the result when fitting a linear function. The shaded histogram
    shows the normalized distribution of $\delta_t$.}
\label{fig:time_resolution}
\end{figure}

The decay-time resolution of the detector directly affects the precision of the
measured CP-violating parameters. Therefore, detailed understanding of the time
resolution is required. The decay-time error that is estimated on an event-by-event
basis during the reconstruction step is underestimated and therefore requires
calibration. In order to perform decay-time
resolution calibration both analyses use a prompt data sample that consists of $J/\psi \to \mu^+ \mu^-$
candidates that originate from the primary vertex and therefore have zero lifetime.
The decay-time distribution of prompt $\bsjpsikk$ candidates is shown in Fig.~\ref{fig:time_resolution}~(a).
Events with negative lifetime are present which can only be possible
due to the resolution effect of the detector. This is used to assess the effective decay-time
resolution of the detector. In order to correct the decay-time uncertainty estimation,
a binned procedure is implemented where the prompt sample is divided in several
subsamples with different values of the per-event error. In each subsample the
effective resolution is assessed by fitting the decay-time distribution. The result
of this procedure is shown in Fig.~\ref{fig:time_resolution}~(b). The relation
between the effective resolution and the per-event error is parametrised with a linear
dependency and the linear calibration parameters are extracted using a $\chi^2$ fit.

\subsection{Reconstruction and selection efficiency}\label{subsec:acceptances}
The geometry of the detector together with the selection requirements cause non-uniform
efficiency as a function of the observables~(three helicity angles and decay-time).
In the case of $\bsjpsipipi$ decay, the invariant mass of two pions is also an
observable in the fit and the efficiency as a function of it is studied separately.
For both decay modes, efficiency as a function of decay angles is evaluated
using simulated samples and then taken into account in the final fit.

The non-uniform shape of the efficiency as a function of decay-time is caused by biasing
selection introduced already in the trigger stage and is described with cubic
splines. Since simulation is not completely reliable in modeling of trigger response,
a data-driven method is used. The control channel $\bdjpsikstar$ is used since
it has a well-known lifetime and kinematics of the decay is similar to the signal channels.
Signal candidates of the control channel are
selected following the selection procedure used for the signal channel. Then the
decay-time acceptance is evaluated and corrected by the ratio of acceptances in
simulated signal and control channels to take into account differences between the
two channels. The final efficiency is represented in the following form:

\begin{equation}
    \varepsilon^{\Bs}_{\mathrm{data}}(t) =
    \varepsilon^{B^{0}}_{\mathrm{data}}(t) \times
    \frac{\varepsilon^{\Bs}_{\mathrm{sim}}(t)}{\varepsilon^{B^{0}}_{\mathrm{sim}}(t)} \,,
\label{eq:timeacc}
\end{equation}

\noindent The differential decay-time rate is then multiplied with evaluated
efficiency as shown in equation~\ref{eq:diff_decay_rate}.

\section{Results}\label{subsec:fit}
\begin{figure}[b]
\centerline{\includegraphics[width=0.9\linewidth]{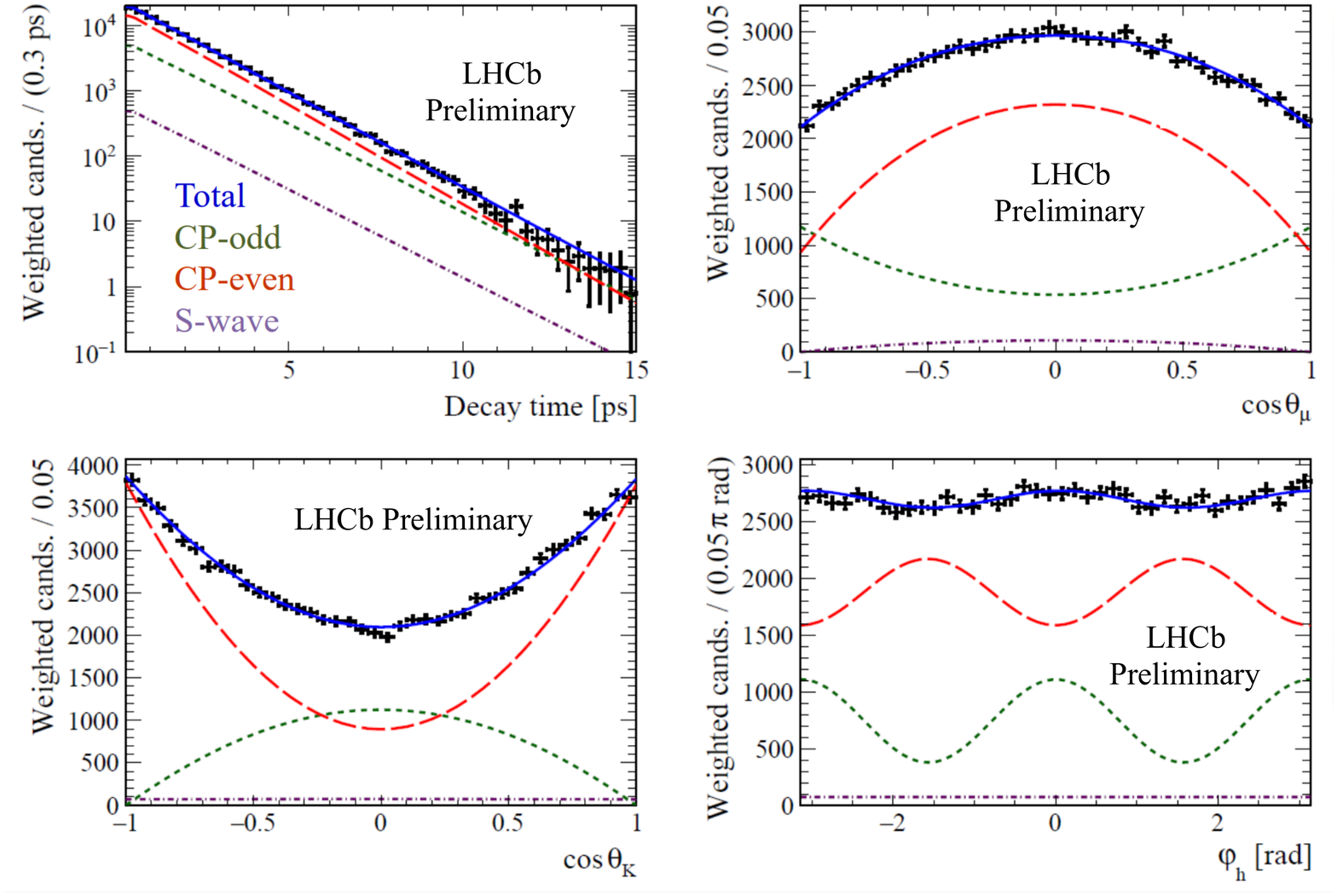}}
\caption{Decay-time and helicity-angle distributions for background subtracted
    $\bsjpsikk$ decays~\protect\cite{kk} with the one-dimensional
    projections of the PDF. The solid blue line shows the total signal contribution,
    which contains CP-even (long-dashed red), CP-odd (short-dashed green) and
    S-wave (dotted-dashed purple) contributions.}
\label{fig:plot_projections1}
\end{figure}

\begin{figure}[h]
\centerline{\includegraphics[width=0.9\linewidth]{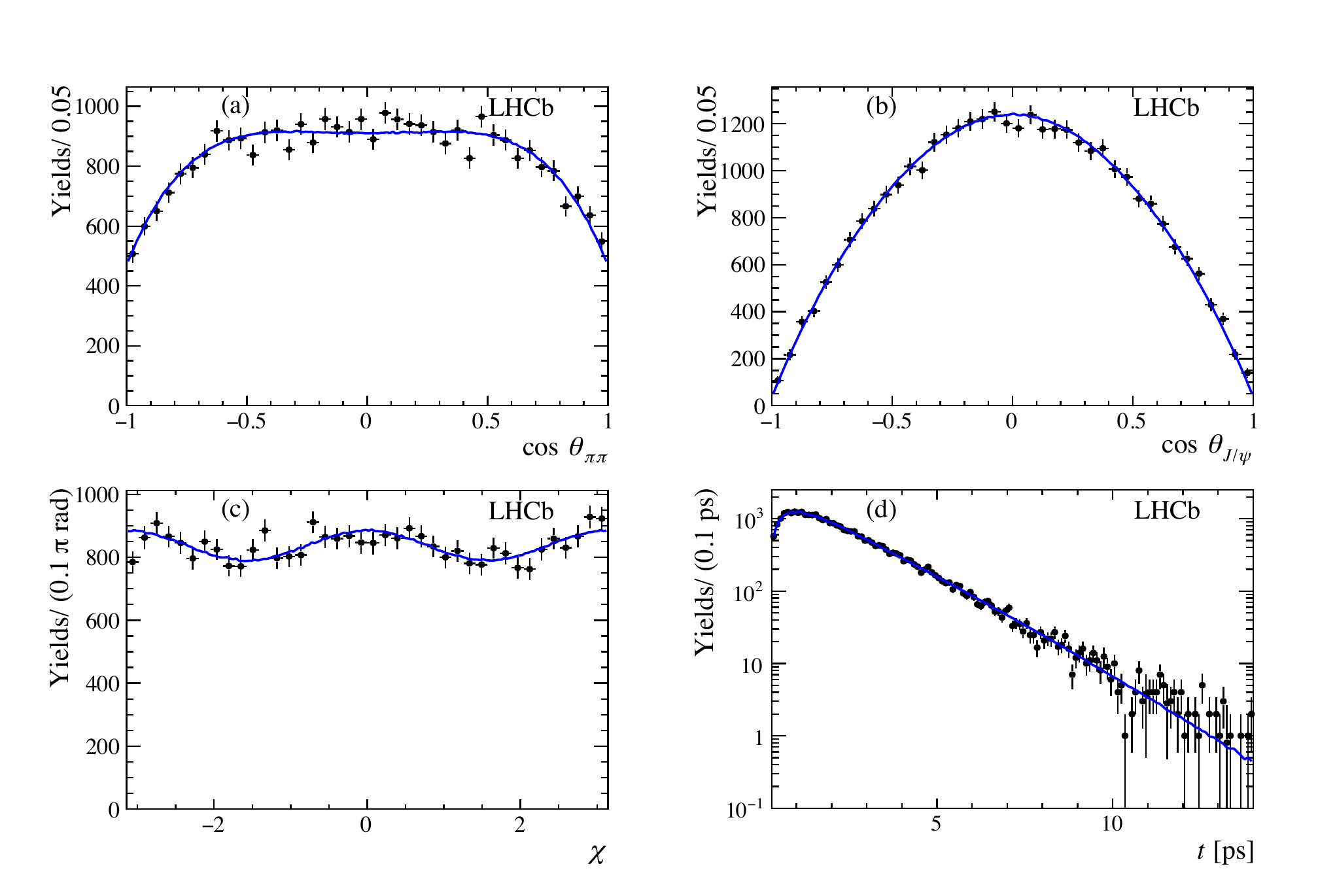}}
\caption{Decay-time and helicity-angle distributions for background subtracted
    $\bsjpsipipi$ decays~\protect\cite{pipi} with the one-dimensional projections of
    the PDF. The solid blue line shows the total signal contribution.}
\label{fig:plot_projections2}
\end{figure}

Taking into account all the inputs described above, a time-dependent maximum
likelihood fit is performed in order to extract the parameters of interest.
The corresponding background-subtracted data distributions with fit projections
for the $\bsjpsikk$ channel are shown in Fig.~\ref{fig:plot_projections1} and
in Fig.~\ref{fig:plot_projections2} for the $\bsjpsipipi$ decay mode.
Table~\ref{tab:results} summarises parameter estimates obtained by the both analyses.

\begin{table}[htb]
\caption[]{Parameters estimates obtained with $\bsjpsikk$ and $\bsjpsipipi$
    decay modes. If two uncertainties are given, the first one is statistical
    and the second one is systematic. If only one is given then it is a
    combination of both statistical and systematic uncertainties calculated
    under the assumption that those contributions are independent. }
\label{tab:results}
\vspace{0.4cm}
\begin{center}
\begin{tabular}{|c|c|c|}
\hline
& $\bsjpsikk$ & $\bsjpsipipi$ \\ \hline
$\phis$, $\mathrm{rad}$ & $-0.080 \pm 0.041 \pm 0.006$ & $-0.057 \pm 0.060 \pm 0.011$ \\
$|\lambda|$ & $1.006 \pm 0.016 \pm 0.006$ & $1.01^{+0.08}_{-0.06} \pm 0.03$ \\
$\Gamma_{s/H}-\Gamma_{B^0}$, ps$^{-1}$ & $-0.0041 \pm 0.0024 \pm 0.0015$ & $-0.050 \pm 0.004 \pm 0.004$  \\
$\dgs$, ps$^{-1}$ & $0.0772 \pm 0.0077 \pm 0.0026$ & - \\ \hline
\end{tabular}
\end{center}
\end{table}

Using a minimum $\chi^2$ fit, the presented results are combined with all the
previous measurements of the $\phis$ phase performed by the LHCb
experiment~\cite{LHCb-PAPER-2014-019,LHCb-PAPER-2016-027,LHCb-PAPER-2014-051,LHCb-PAPER-2017-008}.
The combined values are
\begin{eqnarray}
\label{eq:comb}
\phis &=& -0.040 \pm 0.025 \,\mathrm{rad}\,, \nonumber \\
|\lambda| &=& 0.991 \pm 0.010\,, \nonumber \\
\Gamma_{s}-\Gamma_{B^0} &=& -0.0024 \pm 0.0018\,ps^{-1}\,,\nonumber \\
\Delta\Gamma_{s} &=& 0.0813 \pm 0.0048\,ps^{-1}\,.
\end{eqnarray}
The results are compatible with the SM expectations and with no CPV in the decay
modes under study.


\section*{References}
\begin{thebibliography}{99}

\bibitem{lhcb1}Alves~Jr., A. A. and others, \Journal{JINST}{3}{S08005}{2008}
\bibitem{lhcb2}Aaij, R. and others, \Journal{Int. J. Mod. Phys. A}{30}{1530022}{2015}
\bibitem{CKMfitter}Charles, J. and others, \Journal{\PRD}{84}{033005}{2011}.
\bibitem{Buras:2009if}Buras, Andrzej J., \Journal{PoS}{EPS-HEP2009}{024}{2009}.
\bibitem{Chiang:2009ev}Chiang, Cheng-Wei and Datta, Alakabha and Duraisamy, Murugeswaran and London, David and Nagashima, Makiko and others, \Journal{JHEP}{04}{031}{2010}
\bibitem{pipi}Aaij, Roel and others arXiv {\bf1903.05530} (2019)
\bibitem{kk}Aaij, Roel and others LHCb-PAPER-2019-013
\bibitem{helicity}Aaij, R. and others, \Journal{\PRD}{87}{112010}{2013}
\bibitem{Liu:2013nea}Liu, Xin and Wang, Wei and Xie, Yuehong, \Journal{\PRD}{89}{094010}{2014}.
\bibitem{bdt}Breiman, L. and Friedman, J. H. and Olshen, R. A. and Stone, C. J., Classification and regression trees 1984
\bibitem{splot}Pivk, Muriel and Le Diberder, Francois R., \Journal{\NIMA}{555}{356-369}{2005}
\bibitem{LHCb-PAPER-2014-019}Aaij, R. and others, \Journal{\PLB}{736}{186}{2014}
\bibitem{LHCb-PAPER-2016-027}Aaij, R. and others, \Journal{\PLB}{762}{253}{2016}
\bibitem{LHCb-PAPER-2014-051}Aaij, R. and others, \Journal{\PRL}{113}{211801}{2014}
\bibitem{LHCb-PAPER-2017-008}Aaij, R. and others, \Journal{JHEP}{08}{037}{2017}

\end{thebibliography}
\end{document}